\def\section{\@startsection {section}{1}{\z@}{-26pt plus -1ex minus
    -.2ex}{13pt plus .2ex}{\large \bf}}
\def\subsection{\@startsection{subsection}{2}{\z@}{-13pt plus -1ex minus
   -.2ex}{13pt plus .2ex}{\large \rm}}
\def\subsubsection{\@startsection{subsubsection}{3}{\z@}{-13pt plus
  -1ex minus -.2ex}{-9pt plus .2ex}{\large \em}}
\def\paragraph{\@startsection
     {paragraph}{4}{\z@}{3.25ex plus 1ex minus .2ex}{-1em}{\large\bf}}
\def\subparagraph{\@startsection
     {subparagraph}{4}{\parindent}{3.25ex plus 1ex minus
     .2ex}{-1em}{\large\bf}}

\newcommand{\heading}[1]{\vspace*{15mm}
{\Large\begin{center} {\bf{#1}} \end{center}}}
\renewcommand{\author}[5]{\vspace{5mm}
\begin{center}
{\normalsize \rm #1}\\    
{\normalsize \it #2}\\    
{\normalsize \it #3}\\    
{\normalsize \it #4}\\    
{\normalsize \it #5}\\    
\vspace{0.65cm}\framebox[3.8truecm]{\rule[-1.9cm]{0.cm}{3.8truecm}}
\vspace{0.35cm}\end{center} }

\newcommand{\acknowledgements}[1]{\vspace{7mm} \noindent
	{\normalsize \bf Acknowledgements.\,} {\normalsize #1}}

\def\lsim{~\rlap{$<$}{\lower 1.0ex\hbox{$\sim$}}}
\def\gsim{~\rlap{$>$}{\lower 1.0ex\hbox{$\sim$}}}

\newcommand{\mn}{{\em Mon. Not. R. astr. Soc}}

\newcommand{\physrev}{{\em Phys. Rev.}}
\newcommand{\apj}{{\em Astrophys. J.}}

\renewcommand{\aa}{{\em Astr. Astrophys.}}

\documentstyle[11pt,conf_iap]{article}
\begin{document}
\heading{NON--LINEAR DYNAMICS OF IRROTATIONAL DUST: \\ EULERIAN AND
LAGRANGIAN APPROACHES}

\author{S. MATARRESE$^{1}$, P. CATELAN${^2}$, F. LUCCHIN$^{3}$,
       L. MOSCARDINI$^{3}$, O. PANTANO$^{1}$, D. SAEZ$^{4}$}
       {$^{1}$ Dipartimento di Fisica G. Galilei, Universit\`a di Padova,
       via Marzolo 8, 35131 Padova, Italy.}
       {$^{2}$ International School for Advanced Studies, SISSA, via Beirut 2,
       34014 Trieste, Italy.}
       {$^{3}$ Dipartimento di Astronomia, Universit\`a di Padova,
       vicolo dell'Osservatorio 5, 35122 Padova, Italy.}
       {$^{4}$ Departamento de Fisica Teorica, Universidad de Valencia,
       Burjassot, Valencia, Spain.}

\begin{abstract}{\baselineskip 0.4cm Some recently proposed approximations to
follow the non--linear evolution of collisionless matter perturbations in the
universe are reviewed. The first one, called frozen--flow approximation, is an
Eulerian method within Newtonian theory, and is based on neglecting the role
of particle inertia compared to the damping implied by the Hubble drag. The
second approach is General Relativistic and Lagrangian; it is based on
following the evolution of fluid and geometric observables in the rest frame
of each fluid element, under the only approximation of neglecting the
back--reaction on the system of the gravitational radiation emitted during
non--linear collapse.}
\end{abstract}

\section{Introduction}

A relevant cosmological problem is to understand the physical
processes that occurred during the gravitational collapse of the matter which
gave rise to the observed structure of the universe on large scales. A
complementary issue is to reconstruct the initial conditions of the clustering
process, e.g. the value of the cosmological parameters, the type of dark
matter, the statistics of the primordial perturbations, starting from
observational data such as the spatial distribution of galaxies or their
peculiar velocities. Much work has recently focused in the latter direction,
since more and more data on peculiar velocities of optical galaxies, as well
as very large and complete galaxy redshift surveys have become available both
in the optical and in the infrared.

A widely applied approximation when dealing with the
dynamics of dark matter, either cold or hot, is to treat it as a system of
particles having negligible non--gravitational interactions, a
self--gravitating collisionless system.
The dynamics of such a system is usually approached by different
techniques, depending on the specific application. For instance, the evolution
of small perturbations on a Friedmann--Robertson--Walker (FRW)
background is followed by analytical methods.
The non--linear evolution in cases where some symmetries are present
can also sometimes be followed analytically: typical examples being the
spherical top--hat model for the Newtonian case and the
Tolman--Bondi solution in General Relativity (GR).
There are also useful approximations valid in the
mildly non--linear regime, such as the Zel'dovich
approximation \cite{zel}. Besides this classical approach, a number
of variants have been proposed, all trying to overcome the
inability to follow the development of
structures beyond caustic formation. A promising approach
is adhesion theory (e.g. Ref.\cite{sz} and references therein), where
artificial viscosity is introduced
to mimic the gravitational sticking of particles around pancakes.
An alternative method, called frozen--flow approximation,
recently proposed by Matarrese {\em et al.} \cite{ffa}, allows an
extrapolation beyond the time when orbit--crossing would have occurred
according to the Zel'dovich formulation. This is based on the idea of
extrapolating the growing mode of the linear velocity field beyond its actual
range of validity, while solving exactly the continuity equation.
Other approximations have been explored, such as the second order
Lagrangian perturbation expansion
\cite{bu1,bu2,mou,bou,gra,lac}, and
algorithm proposed by Giavalisco {\em et al.} \cite{gia}.
Different approximations apply in the highly non--linear regime, such as the
hierarchical closure ansatz for the BBGKY equations \cite{p80}.
The most general problem of studying the fully non--linear dynamics
of a collisionless system in Newtonian theory can only be
followed by numerical techniques, such as $N$--body codes.

A General Relativistic (GR) approach to the non--linear evolution of scalar
perturbations of a zero--pressure fluid has been proposed by Matarrese, Pantano
\& Saez \cite{mps}, under the assumption of vanishing vorticity and negligible
gravitational--wave interactions with the rest of the system.
In more technical terms the latter condition amounts to disregarding
the so--called magnetic part of the Weyl tensor, during the non--linear
evolution of perturbations prior to shell--crossing.
Exact GR solutions with vanishing
vorticity and magnetic components have been studied by Barnes \& Rowlingson
\cite{br} in a general context.
The main advantage of such a treatment is that the fluid flow can be entirely
followed in terms of strictly {\em local} evolution equations.
Recently, this kind of approach has attracted some
attention because of the advantages of a purely local algorithm.
In particular, Croudace {\em et al.} \cite {cro} have shown the
connection of the GR pancake solution found in \cite{mps}
with the Szekeres line element \cite{szek}.
Bertschinger \& Jain \cite{bj} have reached relevant conclusions
about the Lagrangian behaviour of fluid elements during collapse.
The relation of the GR treatment with the traditional
Newtonian theory is discussed in detail in Ref.\cite{new}.

\section{Newtonian dynamics in Eulerian form: the frozen--flow approximation}

The standard Newtonian equations for the evolution of collisionless matter
in the universe can be rewritten in terms of suitably rescaled variables and in
comoving coordinates. In particular, it is sometimes convenient to use as time
variable the growth factor of linear perturbations, which in a
flat, matter dominated model, coincides with the expansion factor
$a(t)=a_0(t/t_0)^{2/3}$
(a subscript $0$ will be used for the ``initial time" $t_0$). The Euler
equations read
\begin{equation}
{d {\bf u} \over d a} + {3 \over 2 a} {\bf u} = - {3 \over 2 a}
\nabla \varphi,
\end{equation}
\\
where ${\bf u} \equiv d {\bf x}/ d a$ is a rescaled comoving peculiar
velocity field and the symbol ${d \over d a}$ stands for the total (convective)
derivative ${d \over d a}= {\partial \over \partial a} + {\bf u} \cdot
\nabla$. The continuity equation can be written in terms of
the comoving matter density $\eta({\bf x},t) \equiv \varrho({\bf x}, t)
{}~a^3(t) / \bar \varrho_0 a_0^3$ (where $\bar \varrho_0$ is the mean mass
density at $t_0$)
\begin{equation}
{d \eta \over d a} + \eta \nabla \cdot {\bf u} = 0,
\end{equation}
\\
while the rescaled local gravitational potential
$\varphi \equiv (3t_0^2/2a_0^3) \phi({\bf x},t)$ is
determined by local density inhomogeneities $\delta({\bf x},t) \equiv
\eta({\bf x},t) - 1$ through Poisson's equation
\begin{equation}
\nabla^2 \varphi = {\delta \over a}.
\end{equation}
\\
We restrict our analysis to irrotational flow.

The Zel'dovich approximation, in these variables, corresponds to the
ansatz
${\bf u} = - \nabla \varphi$, as suggested by linear theory.
In this case the Euler and continuity equations decouple from Poisson's one,
and the system describes inertial motion of particles with
initial velocity field impressed by local gravity, as implied by
the growing mode of linear perturbation theory:
${\bf u}_{ZA}({\bf x},\tau) = - \nabla_{\bf q} \varphi_0({\bf q})$, where
${\bf q}$ is the initial (Lagrangian) position and $\tau \equiv a - a_0$.
It follows that particles move along straight trajectories
\begin{equation}
{\bf x}({\bf q},\tau) = {\bf q} - \tau {\bf \nabla}_{\bf q}
\varphi_0({\bf q}).
\end{equation}
\\
The frozen--flow approximation (FFA) can be defined as the
solution of the linearized Euler equations, where in the r.h.s. the growing
mode of the linear gravitational force is assumed,
${\bf u}_{FFA}({\bf x},\tau) = {\bf u}_0({\bf x}) = -
\nabla_{\bf x} \varphi_0({\bf x})$, plus a negligible decaying mode.
In this approximation the peculiar velocity field ${\bf u}({\bf x},a)$ is
{\em frozen} at each point to its initial value, that is
\begin{equation}
{\partial {\bf u} \over \partial \tau} =0,
\end{equation}
\\
which is just the condition for steady flow.
The above equation, together with the continuity
equation, define FFA.
Particle trajectories in FFA are described by the integral
equation
\begin{equation}
\\
{\bf x}({\bf q},\tau) = {\bf q} - \int_0^\tau d \tau' \nabla_{\bf x}
\varphi_0[{\bf x}({\bf q},\tau')].
\end{equation}
\\
Particles update their velocity at each infinitesimal step to the local
value of the linear velocity field, without any
memory of their previous motion, i.e. without {\em inertia}.
Stream--lines are then frozen to their initial shape and multi--stream regions
cannot form. A particle moving according to FFA
has zero component of the velocity in a place where the same
component of the initial gravitational force is zero, it will then
slow down its motion in that direction while approaching
that place. Unlike the Zel'dovich approximation, these particles move along
curved paths: once they come close to a pancake configuration
they curve their trajectories, moving almost parallel to it,
and trying to reach the position of the next filament.
Again they cannot cross it, so they modify their motion,
while approaching it, to finally fall, for $\tau \to \infty$, into
the knots corresponding to the minima of the initial gravitational potential.
This type of dynamics implies an artificial thickening
of particles around pancakes, filaments and knots, which mimics the
gravitational clustering around these structures (though these configurations
do not necessarily occur in the right Eulerian locations, nor they
necessarily involve the right Lagrangian fluid elements).
In assuming that the velocity potential is linearly related at any time to the
local value of the initial gravitational potential, FFA disregards
the non--linear effects caused by the back--reaction of the
evolving mass density on the peculiar velocity field itself (via the
non--linear evolution of the gravitational potential). This implies that
a number of physical processes such as merging of pancakes, fragmentation
and disruption of low--density bridges, are totally absent in the FFA
dynamics. Unlike the velocity field, the FFA density field is
non--locally determined by the initial fluctuations, via
the continuity equation; this is clearly shown by the following analytic
expression
\begin{equation}
1 + \delta_{FFA}({\bf x},\tau) = \exp \int_0^\tau d \tau'
\delta_+[{\bf x}({\bf q},\tau')]
\end{equation}
\\
(where $\delta_+\equiv\delta_0/a_0$).
Brainerd, Scherrer \& Villumsen \cite{lep} have recently shown that a
similar formula also applies if one uses a different approximation
(called LEP: linear evolution of potential), consisting in ``freezing"
the gravitational rather than the velocity potential (see also the
equivalent ``frozen--potential" approach by Bagla \& Padmanabhan \cite{fp}).
This approach shows many features in common to FFA, although multi--stream
regions do occur in this case.

Numerical implementation of FFA is straightforward (for
a more technical discussion, see \cite{ffa}) and involves small
computing time: roughly speaking, FFA consists of a multi--step Zel'dovich
approximation,
and very few steps are required to follow the entire evolution.
Matarrese {\em et al.} \cite{ffa} applied FFA to follow the evolution of
structures in the standard CDM model, and found that it gives a fairly
accurate representation of the density pattern from a resolution scale of
$\sim 500$ km s$^{-1}$, while the two--point correlation function fits quite
well the true non--linear result on even smaller scales.
Further connections of FFA and the Zel'dovich approximation can be found,
based on the Hamilton--Jacobi approach to the non--linear dynamics of
collisionless matter. These, as well as other features of Eulerian
perturbation expansions will be discussed elsewhere \cite{cat}.
Recently, Melott {\em et al.} \cite{mel} have tested FFA vs. the
``truncated" Zel'dovich approximation (e.g. Ref. \cite{tru})
and a full N--body code. They
compare a number of statistics between results of the FFA, truncated
Zel'dovich approximation and N--body simulations and find that FFA performs
reasonably well in a statistical sense, e.g. in reproducing the
counts--in--cell distribution, at small scales, but it does poorly in the
crosscorrelation with N--body, especially in models with high initial
small--scale power.

\section{General Relativistic dynamics in Lagrangian form}

A GR method to follow the development of structures in a collisionless medium
has been proposed in Ref.\cite{mps}; we shall here briefly describe
the basic properties of the method. Further developments of the method are
discussed in \cite{new}.

Let us introduce the equations which govern the dynamics of a
collisionless perfect fluid in GR. A complete treatment of the problem and
a full derivation of the equations presented here can be found in the
review by Ellis \cite{ellis}. We use the signature $(-,+,+,+)$; latin
indices refer to space--time coordinates, $(0,1,2,3)$, greek indices to
spatial ones, $(1,2,3)$. The relativistic dynamics of a collisionless
(i.e. with vanishing pressure) self--gravitating perfect fluid is determined
by Einstein's equations and by the continuity equations for the matter
stress--energy tensor
$T_{ab} = \varrho u_a u_b$, where $\varrho$ is the energy density and $u^a$
the four--velocity of the fluid ($u^a u_a=-1$). It is also useful to define
the spatial projection tensor $h^{ab} \equiv g^{ab} + u^a u^b$
($h_{ab}u^b=0$). Differentiation of the velocity field yields
the tensor $v_{ab} \equiv h_a^{~c} h_b^{~d} u_{c;d}$, for
which $v_{ab}u^b=0$, and the acceleration vector
$\dot u^a \equiv u^a_{~;b} u^b$, which is also space--like,
$\dot u^a u_a=0$. An overdot denotes convective differentiation
with respect to the proper time $t$ of fluid elements, namely
${\dot A}_{a_1 a_2 \cdots a_n} = A_{a_1 a_2 \cdots a_n;b} u^b$.
The antisymmetric part of the tensor $v_{ab}$ is the vorticity tensor
$\omega_{ab} \equiv v_{[ab]}$ (the symbol $_{[..]}$
stands for antisymmetrization and $_{(..)}$ for symmetrization),
describing rigid rotations of fluid elements with respect to
a locally inertial rest frame.
Assuming irrotational motions, $\omega_{ab}=0$, one still has two relevant
quantities: the volume expansion scalar $\Theta \equiv
v^a_{~a}$, and the symmetric and traceless shear tensor, $\sigma_{ab}
\equiv v_{(ab)} - {1 \over 3} \Theta h_{ab}$.
{}From the volume expansion scalar, giving the local rate of isotropic
expansion, one can define a length--scale
$\ell$ through $\Theta=3\dot \ell /\ell$, which reduces
to the scale--factor $a(t)$ in the homogeneous and isotropic
FRW models; in that particular case
$\Theta=3H$, where $H(t)$ is Hubble's constant.
The shear tensor, on the other hand, describes a pure straining in which a
spherical fluid volume is distorted into an ellipsoid with axis lengths
changing at rates determined by the three $\sigma^a_{~b}$ eigenvalues,
$\sigma_1$, $\sigma_2$ and $\sigma_3=-(\sigma_1+\sigma_2)$.
The vanishing trace condition implies
that this deformation leaves the fluid volume invariant, while, in the absence
of vorticity, the principal axes of the shear keep their direction
fixed during the evolution, in a locally inertial rest frame.

The fluid acceleration is only caused by pressure gradients, so in our case
$\dot u^a = 0$: in the absence of pressure each fluid element moves along a
geodesic. The continuity equation reads
\begin{equation}
{\dot \varrho} = - \varrho \Theta.
\end{equation}
\\
The expansion scalar satisfies Raychaudhuri's equation
\begin{equation}
{\dot \Theta} = - {1 \over 3} \Theta^2 - 2 \sigma^2
- 4 \pi G\varrho,
\end{equation}
\\
where $G$ is Newton's constant and $\sigma^2 \equiv {1 \over 2} \sigma^{ab}
\sigma_{ab}$. In the FRW case, $\sigma_{ab}=0$
and the latter equation reduces to the familiar Friedmann one,
$3(\dot H + H^2) = -4 \pi G \varrho$.
The shear is determined by the evolution equation
\begin{equation}
{\dot \sigma}_{ab} = - \sigma_{ac} \sigma^c_{~b} +
{2 \over 3} h_{ab} \sigma^2 - {2 \over 3} \Theta \sigma_{ab} - E_{ab},
\end{equation}
\\
where $E_{ac}\equiv C_{abcd}u^bu^d$ is the electric part of
the Weyl tensor $C_{abcd}$ (the latter being the part of the Riemann curvature
not determined by local sources); $E^a_{~b}$ is also
called the tidal force field, for it contains that part of the
gravitational field which describes tidal interactions; it is symmetric,
traceless and flow orthogonal, $E_{ab}u^b=0$. Tidal forces
act on the fluid flow by inducing shear distortions.
The tensor $E^a_{~b}$ can be diagonalized by going to its principal axes
(which do not generally coincide with those of the shear tensor), with
eigenvalues $E_1$, $E_2$ and $E_3=-(E_1+E_2)$.

{}From the Weyl tensor one defines its
magnetic part $H_{ac}={1 \over 2} \eta_{ab}^{~~~gh}C_{ghcd}u^bu^d$
($\eta_{abcd}$ is the completely antisymmetric four--index tensor),
which is also symmetric, traceless and flow orthogonal and
contains the part of the gravitational field which describes
gravitational waves. Actually, gravitational waves
are represented by the transverse traceless parts of $E_{ab}$ and $H_{ab}$,
satisfying $h^{bc} E_{ab;c}=0$ and $h^{bc} H_{ab;c}=0$ (e.g. Ref.\cite{bde}).
While the tidal force field has a straightforward Newtonian analogue, which
can be written in terms of derivatives of the gravitational potential,
$H_{ab}$ has no Newtonian counterpart. The important
point is that, while in Newton's theory the gravitational potential
is usually determined through a constraint equation, namely Poisson's one,
in GR both $E_{ab}$ and $H_{ab}$ can be calculated through evolution equations.

A useful approximation \cite{mps} is to neglect the influence
of $H_{ab}$ on the evolution of $E_{ab}$: for
initially scalar perturbations of an irrotational perfect fluid, this
amounts to neglecting the interaction of gravitational waves (tensor modes)
with the system. In such a case the tidal force field evolves according to
\begin{equation}
{\dot E}_{ab}= - h_{ab} \sigma^{cd} E_{cd}- \Theta E_{ab} +
3 E_{c(a} \sigma_{b)}^{~~c} - 4 \pi G \varrho \sigma_{ab}.
\end{equation}

Besides these evolution equations, there are many constraints that our
variables have to satisfy; these will be automatically satisfied at any time
during the evolution if one consistently sets up the initial
conditions, e.g. by building up all initial values within linear--theory.

The above approximation, together with the absence of pressure in the fluid,
implies that no explicit spatial gradients occur in the evolution equations,
provided the equations are evaluated in a comoving gauge.
This is because, for vanishing pressure gradients and zero magnetic component,
no wave--like degrees of freedom are allowed: neither sound waves, nor
gravitational radiation. While then absence of sound waves is (within the
dust fluid model) an exact condition, the absence of gravitational
waves is only approximate. Gravitational radiation is indeed expected
to arise during the non--linear collapse of a general (i.e. triaxial)
fluid element. This is, however, not expected to induce relevant
back--reaction on the collapse of structures of cosmological interest.

Under the above assumptions, GR proves an economic way to account for the
mutual gravitational interactions among different fluid elements without the
need of simultaneously evolving all of them. However, as soon as the first
caustics form, multi--stream regions appear and non--local effects start to
play a relevant role in the subsequent evolution of these regions.
There is a strong similarity between this method and the Zel'dovich
approximation: in both cases the evolution of each fluid element is completely
determined by the local initial conditions and can be independently followed
up to the time when it enters a multi--stream region. However, the
present method is exact (except for having disregarded gravitational waves) in
the most general three--dimensional case, while the Zel'dovich algorithm
is only exact for one--dimensional perturbations.
Because of the choice of fluid variables and reference frame
the method is a Lagrangian one: at the end of the
calculations physical observables are known in the rest frame of each fluid
element. The next step is then to
reconstruct the Eulerian density and peculiar velocity fields on
comoving space--like hypersurfaces. This is indeed
possible by integrating additional first--order equations to follow
the relative displacement of neighbouring elements, which are represented by
infinitesimal space--like vectors $\xi^a$, evolving according to
$\dot \xi^a = {1 \over 3} \Theta ~\xi^a + \sigma^a_{~b} ~\xi^b$ \cite{ellis}.

The equations that govern the dynamics of our system form a set of twelve
coupled partial differential equations involving twelve
independent variables: $\sigma_{ab}$ (five), $E_{ab}$
(five), $\varrho$ and $\Theta$.
This can be however reduced to a set of
six equations for six unknowns by going to the simultaneous
local principal axes of the shear and tidal force field.
Three supplementary equations can be solved
next to compute the components of each vector $\xi^a$
and reconstruct the final grid.
For a cubic grid with $N_g^3$ nodes, one would need to solve $6 N_g^3$
first--order equations to obtain the dynamical variables plus
$3(N_g-1)(N_g^2+N_g+1)$ extra ones (still of first--order) for the
relative position vector components.

A numerical procedure based on this GR method was developed in
Ref.\cite{mps}; its accuracy was tested by integrating the
non--linear evolution of suitable spherical perturbations in an otherwise
spatially flat FRW universe and comparing the results with the exact
Tolman--Bondi solution for the same initial profiles.

An exact solution for exactly planar symmetry was
found in Ref.\cite{mps}, which turns out to be locally identical to the
Zel'dovich pancake solution. More in general, one can show that the first
order solutions of these GR equations coincide with the Zel'dovich
approximation.

\section{Conclusions}

We have considered various approximations designed to follow the
non--linear dynamics of collisionless matter.
The FFA method \cite{ffa} is based on solving exactly
the continuity equation while extrapolating the linear approximation for the
peculiar velocity field beyond its actual range of validity: this
allows to prevent the occurrence of orbit crossing, which represents the main
drawback of the Zel'dovich approximation.
Thanks to this property, one obtains an approximate description of the
Eulerian density field at later times and/or on smaller scales
compared to the Zel'dovich approximation.
An advantage of FFA over more advanced techniques,
such as the adhesion theory or a full $N$--body code, is a strong
reduction in the computational time, without a relevant loss of accuracy.
FFA could be applied to obtain fast simulations of the
evolution of structures in the universe on a wider range of scales and with
a larger number of particles. Moreover, FFA could be a reliable tool
in reconstruction procedures of the initial density field from present day
data. The results of Ref.\cite{mel}, however, seem to indicate that
a number of improvements need to be applied. High frequency smoothing on the
initial fluctuations would reduce the excess small--scale clustering
of FFA. Taking in some account the evolution of the velocity
potential beyond linear theory would avoid the unrealistic persistence
of first generation pancakes in the results of FFA. These points will
be discussed in \cite{cat}.

We then described a GR approach \cite{mps} to the non--linear evolution of a
self--gravitating collisionless fluid up to the epoch of caustic formation.
The method relies on the approximation of neglecting the interaction
of gravitational waves with the rest of the system and
assuming irrotational motions.
Under these assumptions a simple Lagrangian picture is obtained which,
after self--consistent
initial conditions have been assigned to a grid, allows to
follow the evolution of each fluid element separately
in its own local rest frame.
This method has two evident advantages: first, being
Lagrangian, it automatically guarantees enhanced resolution in
regions of higher density, i.e. where it is more needed; second,
it allows to follow each fluid element as being completely independent of the
others, which obviously reduces the amount of computer storage needed to evolve
the system. Only at the initial time the conditions ought to be specified
simultaneously on the whole grid, which requires an
amount of computer memory comparable to the one required to construct the
initial conditions in an $N$--body code. For instance, if one considers an
initial random (e.g. Gaussian)
realization of the peculiar gravitational potential in momentum space,
$\hat \varphi({\bf k})$ (e.g. for a given choice of power--spectrum),
initial conditions on the grid should be obtained by inverse Fourier
transforming $k_\alpha k_\beta \hat \varphi({\bf k})$.
An advantage of this method is that it allows to concentrate on
the evolution of a given region, where structures are forming, while its
gravitational interaction with the rest of the universe has been already
included by the set up of the initial conditions.

It would be important to understand how to follow the evolution of the system
in the multi--stream regions which occur after
caustic formation. Several methods have been developed
to follow the evolution in the early non--linear phases or to reconstruct
the initial conditions of the clustering process, most of them trying
to circumvent caustic formation: this can be obtained either by suitable
smoothing procedures \cite{pot}, or by artificially sticking the
particles as their orbits
first cross \cite{sz}, or else by asymptotically slowing down
particle motions, so that orbit mixing never occurs \cite{ffa}.
Finally, one could consider the possibility of replacing the {\em fluid}
approximation with the more realistic picture of a system of non--interacting
particles.

\acknowledgements{
We would like to thank E. Bertschinger, M. Bruni, B. Jain, J. Ostriker,
J. Peebles and G. Tormen for useful discussions.
This work has been partially supported by MURST.}

\vfill
\end{document}